\documentclass[12pt]{article}

\usepackage{amsmath, amsfonts, amssymb, amsthm, natbib}
\usepackage{graphicx,psfrag,epsf}
\usepackage{latexsym}
\usepackage{calc}
\usepackage[ragged, symbol]{footmisc}
\usepackage{grffile}
\usepackage{booktabs}
\usepackage{algorithmic}
\usepackage{algorithm}
\usepackage{caption}
\usepackage{comment}
\usepackage{color,url}
\usepackage{enumitem}
\usepackage{bbm}

\usepackage{bm}

\usepackage[utf8]{inputenc} 
\usepackage[T1]{fontenc}    
 
\usepackage{url}            
\usepackage{nicefrac}       
\usepackage{microtype}      

\usepackage{txfonts}
\usepackage{dsfont}

\usepackage{booktabs} 
\usepackage{amsmath,graphicx}
\usepackage{hyperref}
\usepackage{amsfonts, amssymb, amsthm, natbib}
\usepackage{float}
\usepackage{algorithm,amsmath}
\usepackage{multirow}
\usepackage{savesym}

\begin{document}

\def\spacingset#1{\renewcommand{\baselinestretch}%
{#1}\small\normalsize} \spacingset{1}

%
\title{Automatic Hyperparameter Tuning Method for Local Outlier Factor, with Applications to Anomaly Detection}

%

 \author{Zekun Xu\thanks{
SAS Institute Inc. (Email: \textit{zekun.xu@sas.com})}
 \and
Deovrat Kakde\thanks{
SAS Institute Inc. (Email: \textit{dev.kakde@sas.com})}
 \and
 Arin Chaudhuri\thanks{
 SAS Institute Inc. (Email: \textit{arin.chaudhuri@sas.com})}}

 \maketitle

%
 
%
\begin{abstract}
In recent years, there have been many practical applications of anomaly detection
such as in predictive maintenance, detection of credit fraud, network intrusion, and system failure.
The goal of anomaly detection is to identify in the test data anomalous behaviors that are either rare or unseen in the training data. 
This is a common goal in predictive maintenance, which aims to forecast the imminent faults of an appliance
given abundant samples of normal behaviors.  
Local outlier factor (LOF) is one of the state-of-the-art models used for anomaly detection,
but the predictive performance of LOF depends greatly on the selection of hyperparameters.
In this paper, we propose a novel, heuristic methodology to tune the hyperparameters in LOF.
A tuned LOF model that uses the proposed method shows good predictive performance in both simulations and real data sets.
\end{abstract}


%

\section{Introduction}

Anomaly detection has practical importance in a variety of applications such as predictive maintenance, 
intrusion detection in electronic systems \citep[][]{patcha2007overview,jyothsna2011review},
faults in industrial systems \citep[][]{wise1999comparison}, and medical diagnosis \citep[][]{tarassenko1995novelty,quinn2007known,clifton2011identification}. 
Predictive maintenance setups usually assume that the normal class of data points is well sampled in the training data 
whereas the anomaly class is rare and underrepresented.
This assumption is relevant because large critical systems usually produce abundant data for normal 
activities, but it is the anomalous behaviors (which are scarce and evolving) that can be used to proactively forecast imminent failures
Thus, the challenge in anomaly detection is to be able to identify new types of anomalies in the test data
that are rare or unseen in the available training data.

\begin{figure}[H]
\centering
\includegraphics[scale=0.7]{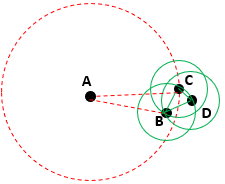}
\caption{A simple example of LOF. Let neighborhood size be 2 and contamination be 0.25.
Point A is identified as an anomaly because it is the most isolated in terms of two nearest neighbors
among the four points.}\label{lofintro}
\end{figure}

Local outlier factor \citep[][]{breunig2000lof} is one of the common methodologies used for anomaly detection,
which has seen many recent applications including credit card fraud detection \citep[][]{chen2007empirical},
system intrusion detection \citep[][]{alshawabkeh2010accelerating}, 
out-of-control detection in freight logistics \citep[][]{ning2012density},
and battery defect diagnosis \citep[][]{zhao2017fault}.
LOF computes an anomaly score by using the local density of each sample point with respect to the 
points in its surrounding neighborhood. 
The local density is inversely correlated with the average distance from a point to its nearest neighbors. 
The anomaly score in LOF is known as the local outlier factor score; its denominator is
the local density of a sample point and its numerator is the average local density of 
the nearest neighbors of that sample point.
 LOF assumes that anomalies are more isolated than normal data points such that
anomalies have a lower local density, or equivalently, a higher local outlier factor score.
LOF uses two hyperparameters: neighborhood size and contamination.
The contamination determines the proportion of the most isolated points 
(points that have the highest local outlier factor scores) to be predicted as anomalies.
Figure \ref{lofintro} presents a simple example of LOF, where we set neighborhood size to be 2
and contamination to be 0.25. Since A is the most isolated point in terms of finding the two nearest neighbors
among the four points, the LOF method predicts it as an anomaly.
In their original LOF paper, Breunig et al. (2000) proposed some guidelines for determining a range for the neighborhood size.
In principle, the number of neighbors should be lower-bounded by the minimum number of points in a cluster
and upper-bounded by the maximum number of  nearest points that can potentially be anomalies.
However, such information is generally not available. Even if such information is available, the optimal 
neighborhood size between the lower bound and upper bound is still undefined.
A second hyperparameter in the LOF algorithm is the contamination, which specifies the proportion
of data points in the training set to be predicted as anomalies. 
The contamination has to be strictly positive in order to form the decision boundaries in LOF. 
In an extreme but not uncommon setting of anomaly detection, there can be zero anomalies in the training data.
In this case, an arbitrary, small threshold has to be chosen for the contamination.\\
These two hyperparameters are critical to the predictive performance in LOF; however,
to the best of our knowledge, no literature has yet focused on tuning both contamination and neighborhood size 
in LOF for anomaly detection.
Since the type and proportion of the anomaly class can be very different between training and testing,
the state-of-the-art K-fold cross validation classification error (or accuracy) does not apply in this setting.
Therefore, in this paper we propose a novel, heuristic strategy for jointly tuning the hyperparameters
in LOF for anomaly detection, and we evaluate this strategy's performance on both moderate and large
data sets in various settings. In addition, we compare the empirical results on real data sets with other benchmark anomaly
detection methods, including one-class SUM \citep[][]{scholkopf2001estimating}
 and isolation forest \citep[][]{liu2008isolation}.

\section{Related Work}

There have been many variants of LOF in the recent years. 
Local correlation integral (Loci) proposed by Papadimitriou et. al (2003), provides an automatic, data-driven approach for
outlier detection that is based on probabilistic reasoning.
Local outlier probability (LoOP) \citep[][]{kriegel2009loop,kriegel2011interpreting} proposes a normalization of the LOF scores to the interval [0,1] 
by using statistical scaling to increase usability across different data sets.
Incremental and memory-efficient LOF methods \citep[][]{pokrajac2007incremental,salehi2016fast} were developed 
so as to efficiently fit an online LOF algorithm in the data stream. 
To make LOF feasible in high-dimensional setting, random projection
 is a common preprocessing step for dimension reduction; it is based on the  Johnson-Lindenstrauss lemma  \citep[][]{dasgupta2000experiments,bingham2001random}. 
Projection-based approximate nearest neighbor methods \citep[][]{liu2005investigation,jones2011randomized}
and approximate LOF methods \citep[][]{lazarevic2005feature,aggarwal2001outlier,de2010finding} have been proposed and evaluated 
in recent literature.

\section{Methodology}

In this paper, we propose a heuristic method to tune the LOF for anomaly detection.
LOF uses two hyperparameters: the first is neighborhood size ($k$), which defines the neighborhood for the computation of local density;
the second is contamination ($c$), which specifies the proportion of points to be labeled as anomalies.
In other words, $k$ determines the score for ranking the training data,
whereas $c$ determines the cutoff position for anomalies.
Let $X\in\mathbb{R}^{n\times p}$ be the training data with a collection of $n$ data points,  
$x_i\in\mathbb{R}^p$.
If $p$ is large, dimension-reduction methods should be used to preprocess the training data
and project them onto a lower-dimensional subspace. 
In predictive maintenance, the anomaly proportion in the training data is usually low
as opposed to the test data, which might contain unseen types of anomalies.
If the anomaly proportion in the training data is known, we can use that as the value for $c$
and tune only the neighborhood size $k$; otherwise, both $k$ and $c$ would have to be tuned in LOF,
which commonly is the case. We assume that anomalies have a lower local relative density 
as compared to normal points, so the top $\lfloor cn \rfloor$ points with
the lowest local density (highest local outlier factor scores) are predicted as anomalies. 

To jointly tune $k$ and $c$, we first define a grid of values for $k$ and $c$,
and compute the local outlier factor score for each training data point under different settings of $k$ and $c$.
For each pair of $k$ and $c$, let $M_{c,k,out}$ and $V_{c,k,out}$ denote the sample mean 
and variance, respectively, of the natural logarithm of local outlier factor scores for
the $\lfloor cn \rfloor$ predicted anomalies (outliers). 
Accordingly, $M_{c,k,in}$ and $V_{c,k,in}$ denote the sample mean and variance, respectively, of the $\log$ local outlier factor scores for
the top  $\lfloor cn \rfloor$ predicted normal points (inliers), which have the highest local outlier factor scores. 
For each pair of $c$ and $k$, we define the standardized difference in the mean log local outlier factor scores between the
predicted anomalies and normal points as
\begin{equation*}
T_{c,k} = \frac{M_{c,k,out}-M_{c,k,in}}{\sqrt{\frac{1}{\lfloor cn \rfloor} \left( V_{c,k,out} + V_{c,k,in}  \right)  }}.
\end{equation*} 
This formulation is similar to that of the classic two-sample $t$-test statistic.
The optimal $k$ for each fixed $c$ is defined as  
$k_{c,opt}=\arg\max_kT_{c,k}$.
If $c$ is known a priori, we only need to find the $k_{c,opt}$ that maximizes the standardized difference 
between outliers and inliers for that $c$.
A logarithm transformation serves to symmetrize the distribution of local outlier factor scores and alleviate the influence of extreme values.
Instead of focusing on all predicted normal points, we focus only on
those $\lfloor cn \rfloor$ normal points that are most similar to the predicted anomalies in terms of their local outlier factor scores.
The intuition behind our focus mimics the idea of support vector machine \citep[][]{cortes1995support} in that
we want to maximize the difference between the predicted anomalies and the normal points 
that are close to the decision boundary.

\begin{figure}[H]
\centering
\includegraphics[scale=0.5]{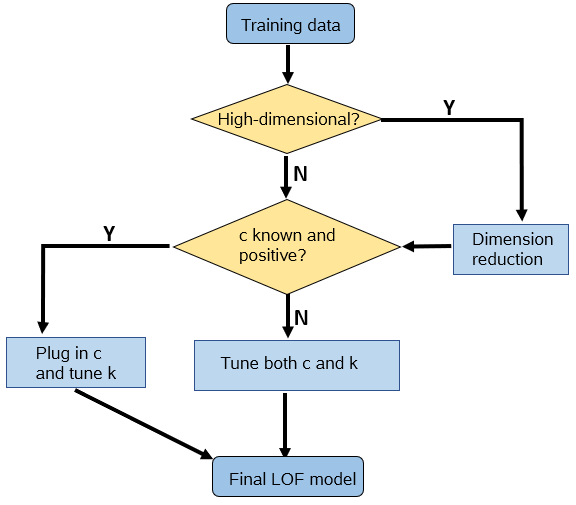}
\caption{Flowchart of training a tuned LOF model.}\label{flowchart}
\end{figure}

We then consider the case when $c$ is not known a priori.
Suppose that for each $c$, the log local outlier factor scores for outliers form a random sample of Gaussian distribution with
mean $\mu_{c,out}$ and variance $\sigma^2_{c,out}$, 
and that the log local outlier factor scores for inliers form a random sample of Gaussian distribution with
mean $\mu_{c,in}$ and variance $\sigma^2_{c,in}$.
Then given $c$, $T_{c,k}$ approximately follows a noncentral $t$ distribution with $2\lfloor cn \rfloor-2$
degrees of freedom and noncentrality parameter 
$\frac{\mu_{c,out}-\mu_{c,in}}{\sqrt{\frac{1}{\lfloor cn \rfloor} \left( \sigma^2_{c,out} + \sigma^2_{c,in}  \right)  }}$.
We cannot directly compare the largest standardized difference $T_{c,k_{c,opt}}$ across different
values of $c$ because $T_{c,k}$ follows different noncentral $t$ distributions depending on $c$.
Instead, we can compare the quantiles that correspond to $T_{c,k_{c,opt}}$ in each respective noncentral distribution
so that the comparison is on the same scale.
Define $c_{opt} = \arg\max_cP(Z<T_{c,k_{c,opt}}; \mathit{df}_c ,\mathit{ncp}_c)$, where the random variable $Z$
follows a noncentral $t$ distribution with $\mathit{df}_c$ degrees of freedom and $\mathit{ncp}_c$ noncentrality parameter.
Thus, the optimal $c$ is the one where $T_{c,k_{c,opt}}$ is the largest quantile in the corresponding $t$ distribution
as compared to the others. Since we do not observe the noncentrality parameter, it will be
estimated by plugging in sample means and variances for the true population counterparts.
Figure \ref{flowchart} displays the flowchart of procedures for training a tuned LOF model.

 \begin{algorithm}
    \caption{Tuning algorithm for LOF}
  \begin{algorithmic}[1]
\STATE training data $X\in\mathbb{R}^{n\times p}$
\STATE a grid of feasible values $\textrm{grid}_c$ for contamination $c$
 \STATE  a grid of feasible values $\textrm{grid}_k$ for neighborhood size $k$
    \FOR{each $c\in \textrm{grid}_c$}
     \FOR{each $k\in \textrm{grid}_k$}
	 \STATE set $M_{c,k,out}$ to be mean log LOF for the ${\lfloor cn \rfloor}$ outliers
 	\STATE set $M_{c,k,in}$ to be mean log LOF for the ${\lfloor cn \rfloor}$  inliers
 	\STATE set $V_{c,k,out}$ to be variance of  log LOF for the ${\lfloor cn \rfloor}$ outliers
 	\STATE set $V_{c,k,in}$ to be variance of  log LOF for the ${\lfloor cn \rfloor}$  inliers
         \STATE set
              $T_{c,k} = \frac{M_{c,k,out}-M_{c,k,in}}{\sqrt{\frac{1}{\lfloor cn \rfloor} \left( V_{c,k,out} + V_{c,k,in}  \right)  }}$
\ENDFOR
    \STATE set $M_{c,out}$ to be mean $M_{c,k,out}$ over $k\in \textrm{grid}_k$
 \STATE set $M_{c,in}$ to be mean $M_{c,k,in}$ over $k\in \textrm{grid}_k$
 \STATE set $V_{c,out}$ to be mean $V_{c,k,out}$ over $k\in \textrm{grid}_k$
 \STATE set $V_{c,in}$ to be mean $V_{c,k,in}$ over $k\in \textrm{grid}_k$
    \STATE set $\textit{ncp}_c=\frac{M_{c,out}-M_{c,in}}{\sqrt{\frac{1}{\lfloor cn \rfloor} \left( V_{c,out} + V_{c,in}  \right)  }}$
     \STATE set $\textit{df}_c=2\lfloor cn \rfloor-2$
    \STATE set $k_{c,opt}=\arg\max_kT_{c,k}$
    \ENDFOR
    \STATE set $c_{opt}=\arg\max_cP(Z<T_{c,k_{c,opt}}; df_c,ncp_c)$, where the random variable $Z$
follows a noncentral $t$ distribution with $\textit{df}_c$ degrees of freedom and $\textit{ncp}_c$ noncentrality parameter
  \end{algorithmic}
\end{algorithm}

\section{Experimental Results}

\subsection{Performance measures}

We use both the area under the  ROC curve (AUC) and the F1 score to evaluate the goodness of 
the optimal parameters that are tuned by the proposed metric.
The F1 score is defined as   
\begin{equation*}
F1 = \frac{2\times \textrm{precision}\times \textrm{recall}}{\textrm{precision+recall}}.
\end{equation*}
The F1 score is a measure of precision and recall at a particular threshold value on the ROC curve,
and AUC is an average over all the threshold values.
 

\subsection{Evaluations on small data sets}

We first assess the performance of the proposed tuning metric on three small data sets by checking 
how the selected optimal neighborhood size and contamination perform in terms of the AUC and F1 score.
Since the data dimension is low, no dimension reduction is needed in the data preprocessing.

\begin{figure}[H]
\centering
\includegraphics[scale=0.7]{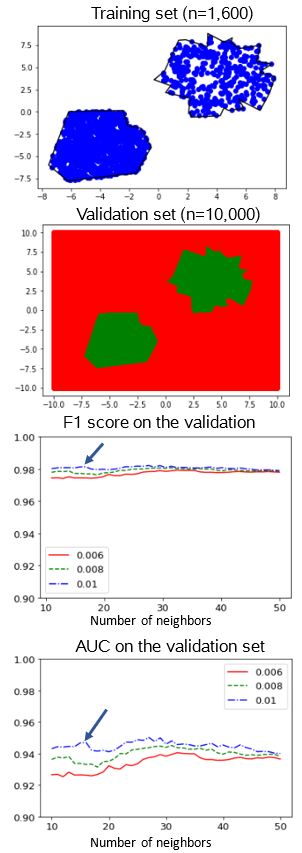}
\caption{The first plot shows the training data. 
The second plot shows the 2-D grid of validation data.
The third and fourth plots display the F1 score and AUC, respectively, on the validation set for different parameter values.
The arrows point to the parameters that were selected using the proposed tuning metric,
where the selected contamination is 0.01 and the neighborhood size is 16.
The F1 score and AUC at the tuned parameter settings are close to the optimal values on
the prespecified grids.}\label{polygons1}
\end{figure}

{\bf{Polygons data:}} This synthetic training set contains 1,600 points, which are uniformly sampled within a mixture of two randomly generated polygons
as shown in Figure \ref{polygons1},
where one polygon has a higher density than the other. 
Since no points are sampled outside the boundaries of the polygons,
the anomaly proportion is 0 in the training set.
The 10,000 data points in the synthetic validation set form a dense two-dimensional (2-D) mesh grid with both axes ranging 
from --10 to 10. The points inside the true boundaries are labeled as normal; the points outside are labeled anomalies. 

\begin{figure}[H]
\centering
\includegraphics[scale=0.73]{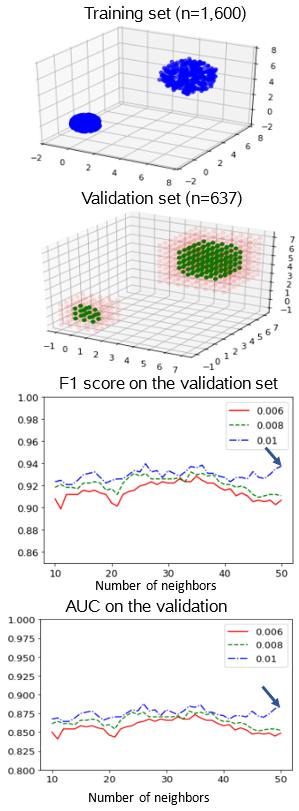}
\caption{The first plot shows the training data. 
The second plot shows the 3-D grid of validation data.
The third and fourth plots display the F1 score and AUC, respectively, on the validation set for different parameter values.
The arrows point to the parameters that were selected using the proposed tuning metric, 
where the selected contamination is 0.01 and the neighborhood size is 48.
The F1 score and AUC at the tuned parameter settings are close to the optimal values on
the prespecified grids.}\label{ball1}
\end{figure}

{\bf{Balls data:}} This synthetic training set contains 1,600 points, which are uniformly sampled within a mixture of two three-dimensional (3-D) balls
as shown in Figure \ref{ball1}, 
where the ball centered at the origin has a smaller radius than the ball centered at (5,5,5). 
Since no points are sampled outside the boundary of the balls,
the anomaly proportion is 0 in the training set.
The 637  points in the synthetic validation set form two 3-D cubes, with each cube enveloping one of the training balls. 
 The points inside the true boundaries are labeled as normal; the points outside are labeled anomalies. 

{\bf{Metal data:}} This engineering data set is used in \cite{wise1999comparison}; it consists of the eight engineering variables from a LAM 9600 metal etcher over the course of etching 129 wafers (108 normal wafers and 21 wafers in which faults were intentionally induced during the same experiments). 
In the training set, we include 90\% of the normal wafers data. The validation set is the entire data set.

\begin{table}[H]
\small
\centering
\begin{tabular}{llcc}
\hline
Name & $p$ & $n$ (Training) & Anomaly$/n$ (Validation) \\
\hline
Polygons  & 2 & $1,600$   & $2,221/10,000$ (22\%) \\
Balls  & 3 & $1,600$ & $98/637$ (15\%)  \\
Metal & 8 & $95$ & $21/129$ (16\%)  \\
\hline      
\end{tabular}
\caption{List of small data sets.}
\label{listsmall}
\end{table}

For both the polygons data and the balls data, the grid of values for neighborhood ranges from 10 to 50 incrementing  by 1,
and the three contamination levels considered are 0.006, 0.008, and 0.01. 
In the metal data,
the grid for neighborhood ranges from 10 to 25 incrementing by 1, and the three contamination levels considered
are 0.08, 0.1, and 0.12.
Table \ref{tablesmall} shows the results on the three small data sets,
where the  proposed method produces a tuned LOF that has both F1 score and AUC
very close to the optimal upper bound values on the prespecifed grids.
 
\begin{table}[H]
\small
\centering
\begin{tabular}{lllllll}
\hline
\multicolumn{1}{l}{Data} & \multicolumn{1}{l}{Tuned $c$} & \multicolumn{1}{l}{Tuned $k$} & \multicolumn{2}{c}{F1} & \multicolumn{2}{c}{AUC} \\
\cline{4-7}
\multicolumn{1}{l}{} &\multicolumn{1}{l}{}&\multicolumn{1}{l}{}&\multicolumn{1}{l}{Tuned}&\multicolumn{1}{l}{Best}&\multicolumn{1}{l}{Tuned}&\multicolumn{1}{l}{Best}\\
\hline
Polygons  & 0.01 & 16  & 0.981 & 0.982 & 0.947 & 0.950\\
Balls  & 0.01 & 48 & 0.930 & 0.939 & 0.875 & 0.888\\
Metal & 0.10 & 14 & 0.844 & 0.844 & 0.886 & 0.886  \\
\hline      
\end{tabular}
\caption{Performance of tuned LOF on the three small data sets. The F1 score and the AUC from the model tuned by using the proposed method are
very close to the optimal values on the prespecified grids.}
\label{tablesmall}
\end{table}

 \begin{figure}[H]
\centering
\includegraphics[scale=0.8]{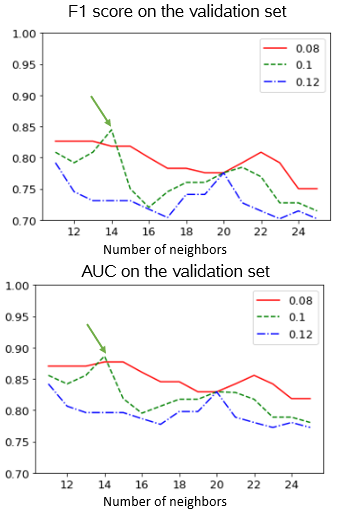}
\caption{The two plots show the F1 score and AUC, respectively, on the validation set for different parameter values.
The arrows point to the parameters that were selected by using the proposed tuning metric.
The selected contamination is 0.1, and the neighborhood size is 14.
The F1 score and the AUC at the tuned parameter setting agree well with the actual peak positions.}\label{metal}
\end{figure}

\subsection{Evaluations on large data sets}

To evaluate the performance of the proposed tuning metric on large data sets, Gaussian random projection is
implemented as a preprocessing step for dimension reduction. 
We do not discuss how to choose the dimension of the projected subspace, because 
dimension reduction is only for the purpose of computation feasibility in this paper.
The computation cost of LOF is $\textit{np}$ times the cost of a $k$-nearest-neighbor (KNN) query,
which is needed in searching the neighborhood for each sample point. 
For low-dimensional data, a grid-based approach can be used
to search for nearest neighbors so that the KNN query is constant in $n$. 
For high-dimensional data, the KNN query on average takes $O(\log n)$, with the worst case of $O(n)$,
which would make the LOF algorithm extremely slow for large, high-dimensional data.
In this paper, we use random projection for dimension reduction to make the computation feasible
for the repetitive running of the LOF algorithm on large data sets. 
In practice, we recommend that the dimension of the data be reduced
to the largest subspace that the computing resources can handle.\\
\par We assessed performance of the LOF method on the following data sets:\\
\par {\bf{Spheres}}$\times${\bf{100:}} We generated 100 mixtures of 100-dimensional spheres data.
In each mixture, the training set contains 100,000 points uniformly sampled from 
a random number (between 2 and 10) of spheres.
Since no points are sampled outside the boundary of the spheres,
the anomaly proportion is 0 in the training set.
For the validation set in each mixture, 10,000 points are randomly sampled around
each of the training spheres with 0.05 probability of being outside the boundaries (anomalies).

{\bf{Cubes}}$\times${\bf{100:}} We generated 100 mixtures of 100-dimensional cubes data.
In each mixture, the training set contains 100,000 points uniformly sampled from 
a random number (between 2 and 10) of cubes with dimension equal to 100.
Since no points are sampled outside the boundary of the cubes,
the anomaly proportion is 0 in the training set.
For the validation set in each mixture, 10,000 points are randomly sampled around
each of the training cubes with 0.05 probability of being outside the boundaries (anomalies).

{\bf{Smtp:}} This data set is a subset from the original KDD Cup 1999 data set from the UCI Machine Learning Repository \citep[][]{hettich1999uci},
where the service attribute is smtp.
The training set consists of 9,598 samples of normal internet connections and 36 continuous variables.
The validation set contains 1,183 anomalies out of 96,554 samples (1.2\%).

{\bf{Http:}}  This data set is also a subset from the original KDD Cup 1999 data set from UCI Machine Learning Repository \citep[][]{hettich1999uci},
where the service attribute is http.
The training set consists of 61,886 samples of normal internet connections and 36 continuous variables.
The validation set contains 4,045 anomalies out of 623,091 samples (0.6\%).

{\bf{Credit:}} This credit card fraud detection data set has been collected during a research collaboration
of Worldline and the Machine Learning Group of Universit\'e Libre de Bruxelles \citep[][]{dal2015calibrating}, 
which contains 284,807 records and 28 continuous variables.
The training set consists of 142,157 normal credit card activity records.
The validation set contains 492 fraudulent activity records out of 284,807 samples (0.2\%).

{\bf{Mnist:}} This data set is a subset from the publicly available MNIST database of handwritten digits \citep[][]{lecun1998gradient}.
The training set consists of 12,665 samples for digits \textquotedblleft{0}\textquotedblright and \textquotedblleft{1}\textquotedblright, which are defined as normal data in this specific application.
The validation set consists of 10,000 samples for all 10 digits, where there are 7,885 (78.9\%) anomalies.

 \begin{table}[H]
\small
\centering
\begin{tabular}{lrrc}
\hline
Name  & $p$ & $n$ (Training) & Anomaly$/n$ (Validation) \\
\hline
Spheres$\times 100$  & 100 & 100,000 & $5,000/100,000$ (5\%) \\
Cubes$\times 100$  & 100 & 100,000 & $5,000/100,000$ (5\%) \\
Smtp  & 36 & 9,598 & $1,183/96,554$ (1.2\%) \\ 
Http  & 36 & 61,886 &  $4,045/623,091$ (0.6\%) \\
Credit & 28 & 142,157 & 492/284,807 (0.2\%)\\
Mnist & 784 & 12,665 & $7,885/10,000$ (78.9\%) \\
\hline      
\end{tabular}
\caption{List of large data sets.}
\label{listlarge}
\end{table}


Table \ref{resultlarge1} shows the performance of the tuning metric on the synthetic Cubes$\times100$ and Spheres$\times100$ data.
After tuning, the mean F1 score and AUC after tuning are high and approach the best upper bound values in both cases,
indicating good predictive performance of the tuned parameter settings.
 For the reduced subspace dimension of 3 with sample size 100,000,
the average running time for LOF in both cases is smaller than 6 seconds,
which shows the scalability of the tuning algorithm for a large sample size.
Table \ref{resultlarge2} compares the tuned LOF versus other benchmark anomaly detection methods (one-class SVM and isolation forest)
on large real data sets. For the first three data sets (Http, Smtp, and Credit), Gaussian random projection is used to reduce the dimension to 3.
For the Mnist data, the reduced subspace dimension is 10 because the original data is high-dimensional. 
We repeat the random projection process 10 times and compare the mean (standard error) of the F1 score and the AUC between different methods. 
LOF is tuned using the proposed metric, whereas the hyperparameters in one-class SVM and isolation forest are chosen to be
the configuration that has the highest F1 and AUC on the validation set.
In the Http and Smtp data sets, the performance of the tuned LOF is comparable to the best result from one-class SVM;  
in Credit and Mnist, the tuned LOF has a higher mean F1 score and AUC than the other two benchmark methods.
Note that the F1 scores from all methods are low on the Credit data, which might imply that the anomalies
are not fully identifiable from the normal data in this case. 

\begin{table}[H]
\small
\centering
\begin{tabular}{llllll}
\hline
\multicolumn{1}{l}{\multirow{2}{*}{Data}}  &  \multicolumn{2}{c}{Mean F1} & \multicolumn{2}{c}{Mean  AUC} &\multicolumn{1}{l}{Mean computation} \\
\cline{2-3} \cline{4-5}
\multicolumn{1}{l}{} & \multicolumn{1}{c}{Tuned} & \multicolumn{1}{c}{Best} & \multicolumn{1}{c}{Tuned} & \multicolumn{1}{c}{Best} &\multicolumn{1}{l}{time (sec)}\\
\hline
Spheres$\times$100 & 0.955 & 0.959 & 0.988 & 0.994 & 5.77\\
		    & (0.022)  & (0.022) & (0.006) & (0.002) &  \\
Cubes$\times$100 & 0.937 & 0.976& 0.987 & 0.991 & 5.79 \\
                            & (0.043) & (0.005) & (0.005) & (0.002) &  \\
\hline
\end{tabular}
\caption{Mean (standard error) of F1 score and AUC on the synthetic Cubes$\times$100 and Spheres$\times$100 data. In each of the 100 mixtures, 100,000 points are 
randomly sampled from a mixture of 100-dimensional cubes (spheres). 
In the preprocessing, random projection is used to reduced the dimension to 3. The best upper bounds of the
F1 score and AUC are computed using the maximum F1 score and AUC among the specified grid values in each repetition. The results show that
the mean of F1 score and AUC after tuning are close to the optimal values.}
\label{resultlarge1}
\end{table}

\begin{table}[H]
\small
\centering
\begin{tabular}{lllllll}
\hline
\multicolumn{1}{l}{\multirow{2}{*}{Data}}  &  \multicolumn{3}{c}{Mean F1} & \multicolumn{3}{c}{Mean AUC} \\
\cline{2-4} \cline{5-7} 
\multicolumn{1}{l}{} & \multicolumn{1}{c}{LOF} & \multicolumn{1}{c}{SVM} & \multicolumn{1}{c}{IForest} &
 \multicolumn{1}{c}{LOF} & \multicolumn{1}{c}{SVM} & \multicolumn{1}{c}{IForest}  \\
\hline
Http & 0.558 & {\bf{0.610}} & 0.356 & {\bf{0.849}} & 0.834 & 0.644  \\
       & (0.157)  &  (0.107) & (0.109) & (0.066) &  (0.0575) & (0.043)  \\
Smtp & 0.662 & {\bf{0.687}} & 0.637 & 0.800 & {\bf{0.814}} & 0.745 \\
     & (0.166) & (0.167)  & (0.062)  & (0.057) & (0.058) & (0.030) \\
Credit & {\bf{0.425}} & 0.311 & 0.295 & {\bf{0.762}} & 0.699 & 0.620 \\
              & (0.148) & (0.112) & (0.095) & (0.064) & (0.056) & (0.038)  \\
Mnist & {\bf{0.824}} & 0.522 & 0.570 & {\bf{0.728}} & 0.628 & 0.616  \\ 
   &   (0.053) & (0.056) & (0.048) & (0.036) & (0.011) & (0.013)  \\
\hline
\end{tabular}
\caption{Comparison of mean (standard error) of F1 score and AUC
 among LOF, one-class SVM, and isolation forest after preprocessing by random projection. 
For the first three data sets, random projection is used to reduce the dimension to 3.
For the Mnist data, random projection is used to reduce the dimension to 10 because the original data is high-dimensional. 
LOF is tuned using the proposed standardized difference on the training set.
The F1 score and AUC for SVM and IForest are the best values in the prespecified grids of parameters. 
We repeat the preprocessing of random projection 10 times and report the mean F1 score and AUC
for each method.}
\label{resultlarge2}
\end{table}

\section{Conclusions}

We propose a heuristic methodology for jointly tuning the hyperparameters of contamination and neighborhood size in the LOF algorithm,
and we comprehensively evaluated this methodology on both small and large data sets.
In small data sets, the tuned hyperparameters correspond well to settings that have the highest F1 score and AUC.
In large data sets, Gaussian random projection is used in the preprocessing step for dimension reduction, whose sole purpose
is to improve computation efficiency. The predictive performance of the tuned LOF is comparable to the predictive performance with the best results from one-class SVM on
the Http and Smtp data, and it outperforms all the other methods on Credit and Mnist data.
 
Although the proposed tuning method works reasonably well in general, it is by no means guaranteed
that the tuned parameters will maximize either the F1 score or the AUC. This is exactly the challenge in anomaly detection
where the test data differ from the training in terms of the anomaly type and proportion. In order for the proposed
tuning method to have good performance, we need to assume that the normal data are well sampled in the training data
and that the anomalies can be identified from the normal data in terms of their relative local density.
As long as those assumptions are not severely violated, the proposed metric (which is based on maximizing the standardized $\log(\text{LOF})$ difference)
will manage to arrive at a decent parameter configuration that differentiates the anomalies from the normal data. 
In future work, extending the tuning methodology to the setting of incremental
LOF for streaming data is worth exploring.

\section{Acknowledgments}

Authors would like to thank Anne Baxter, Principal Technical Editor at SAS, for her assistance in creating this
manuscript.
%
\bibliographystyle{Chicago}
\bibliography{sample-bibliography}

%

\end{document}